\documentstyle[twoside]{siamstyl}
\title{
A SHERMAN MORRISON WOODBURY IDENTITY FOR RANK \\
AUGMENTING MATRICES WITH APPLICATION TO CENTERING%
	\thanks{Received by the editors July 12, 1990; accepted for
		publication February 19, 1991.}}
\author{Kurt S. Riedel%
	\thanks{Courant Institute of Mathematical Sciences,
		New York University, New York, New York 10012.
 The work of this author was supported by the
U.S. Department of Energy Grant No. DE-FG02-86ER53223.}
}
\begin{document}
\markboth{\sc Kurt S. Riedel}{\sc 
Sherman Morrison Identity for Rank
Augmenting}
\pagestyle{myheadings}
\setcounter{page}{79}
\maketitle
\vspace*{-1.2in}
\font\sixrm=cmr6
\font\fiverm=cmr5
{\sixrm\baselineskip10pt
\noindent
SIAM J.\ M{\fiverm AT.} A{\fiverm NAL.}\hspace*{\fill}
{\scriptsize{\copyright}} 1991 Society for Industrial and
Applied\vspace*{-1.2mm} Mathematics\\
Vol.\ 12, No.\ 1, pp. 80--95, January 1991 \hspace*{\fill} 000\par}
\normalsize
\vspace*{0.9in}
\def\vec{}
\def\Yu{\bf Y\,}
\def\Xuu{{\bf X}\,}
\def\Siuu{{\bf \Sigma}\, }
\def\Au{\bf A}
\def\alu{\bf\alpha}
\def\thu{\bf{\theta}}
\def\beu{\bf\beta}
\def\Auu{{\bf A}\,}
\def\Lauu{{\bf \Lambda}}
\def\ev{\vec{e}}
\def\xbv{\vec{\bar{x}}}
\def\xv{\vec{x}}
\def\vv{\vec{v}}
\def\wv{\vec{w}}
\def\ssg{\sigma^2}
\def\bvh{\vec{\hat{b}}}
\def\btvh{\hat{\vec{ \beta}}}
\def\btv{\vec{ \beta}}
\def\nl{\hfill\break}       
\def\np{\vfill\eject}       
\def\nsect{\vskip 2pc\noindent}       
\def\nli{\hfill\break\noindent}       
\def\ni{\noindent}
\def\IR{I\kern-.255em R}
\def\eps{\epsilon }
\def\Yu{\bf Y\,}
\def\Xuu{{\bf X}\,}
\def\Puu{{\bf P}\,}
\def\Omuu{{\bf \Omega}\, }
\def\Siuu{{\bf \Sigma}\, }
\def\Suu{{\bf S}\, }
\def\muu{\bf{\mu}\, }
\def\eu{\bf e\,}
\def\Wuud{{\bf W}_d}
\def\Au{\bf A}
\def\Bu{\bf B\,}
\def\Eu{\bf E\,}
\def\bv{\vec \beta}
\def\beu{\bf\beta}
\def\euu{{\bf e}\,}
\def\Auu{{\bf A}\,}
\def\Buu{{\bf B}\,}
\def\Cuu{{\bf C}\,}
\def\Euu{{\bf E}\,}
\def\Iuu{{\bf I}\,}
\def\Guu{{\bf G}\,}
\def\Quu{{\bf Q}\,}
\def\Quuii{{\bf Q}_{(ii)}\,}
\def\Lauu{{\bf \Lambda}}
\def\Lluu{{\bf \Lambda}}
\def\Uuu{{\bf U}\,}
\def\Vuu{{\bf V}\,}
\def\Wuu{{\bf W}\,}
\def\Zuu{{\bf Z}\,}
\def\aluu{{\bf\alpha}}
\def\beuu{{\bf\beta}}
\begin{abstract}
Matrices of the form $\Auu + (\Vuu_1 + \Wuu_1)\Guu(\Vuu_2 + \Wuu_2)^*$ are
considered where $\Auu$ is a $singular$ $\ell \times \ell$ matrix and
$\Guu$ is a nonsingular $k \times  k$ matrix, $k \le \ell$. 
Let the columns of $\Vuu_1$
be in the column space of $\Auu$ and the columns of 
$\Wuu_1$ be orthogonal to $\Auu$.
Similarly, let the columns of $\Vuu_2$
be in the column space of $\Auu^*$ and 
the columns of $\Wuu_2$ be orthogonal to $\Auu^*$.
An explicit expression for the inverse is given, provided that
$\Wuu_i^* \Wuu_i$ has rank $k$.
An application to
centering covariance matrices about the mean is given.

\end{abstract}
\begin{AMSMOS}
65R10, 33A65, 35K05, 62G20, 65P05
\end{AMSMOS}


The wellknown Sherman-Morrison-Woodbury matrix identity [1]:

$$
( \Auu + \Xuu_1 \Guu \Xuu_2^T)^{-1}
= \Auu^{-1} \ - \
 \Auu^{-1} \Xuu_1 ( \Guu^{-1} +  \Xuu_2^T \Auu^{-1} \Xuu_1 )^{-1} 
 \Xuu_2^T \Auu^{-1} 
\eqno (1)
$$
is widely used. 
Several excellent review articles have appeared 
recently [2-4]. However (1) is only valid when 
$\Auu$ is nonsingular
\footnote{ We denote the transpose of a matrix, $\Auu$ by $\Auu^T$
and the hermitian or conjugate transpose by $\Auu^*$.}.
In this article, we consider matrix inverses of the form
$ \Auu + \Xuu_1 \Guu \Xuu_2^T$ where the rank of
$ \Auu + \Xuu_1 \Guu \Xuu_2^T$ is larger than the rank of $\Auu$.

We decompose the matrix $\Xuu_1$ into $\Vuu_1 + \Wuu_1$, where 
the columns of
$\Vuu_1$ are contained in the column space of $\Auu$ and 
the columns of $\Wuu_1$
are orthogonal to it. Similarly, we decompose
$\Xuu_2$ into $\Vuu_2 + \Wuu_2$, where the columns of 
$\Vuu_2$ are contained in the column space of $\Auu^*$ and the
columns of $\Wuu_2$
are orthogonal to $M(\Auu^*)$ .
We denote the column space of $\Auu$ by
$M(\Auu)$. The Moore-Penrose generalized inverse will be denoted by the
superscript $^+$. We denote the $k \times  k$ matrix $\  \Wuu_i^* \Wuu_i$
by $\Buu_i$ and define {$\Cuu_i \equiv \Wuu_i (\Wuu_i^* \Wuu_i)^{-1}$.
We will require $\Buu_i$ to be nonsingular.
However the rank of the perturbation, $k$, can be significantly less
than the size of the original matrix. We note that $\Vuu_i^* \Wuu_i=0$
and $\Wuu_i^* \Cuu_i  = \Iuu_{k}$.
Finally the projection operator onto the column space of $\Wuu$ satisfies
$\Wuu_i\Buu_i^{-1} \Wuu_i^*  = \Wuu_1\Cuu_1^* = \Cuu_2\Wuu_2^*  $.

\vspace{.1in}

{\bf Theorem 1}.  Let $\Auu$ be a $\ell \times \ell$ 
 matrix of rank $\ell_1$, $\ell_1 < \ell$,
 $\Vuu_i$ and $\Wuu_i$ be $\ell \times k$ matrices
and $\Guu$ be a $k \times k$ nonsingular matrix.
Let the columns of $\Vuu_1 \in M(\Auu)$ and 
the columns of $\Wuu_1$ be orthogonal to $M(\Auu)$.
Similarly, let the columns of $\Vuu_2 \in M(\Auu^*)$ and 
the columns of $\Wuu_2$ be orthogonal to $M(\Auu^*)$.
Let $\Buu_i \equiv \Wuu_i^* \Wuu_i$ have rank $k$. 
The matrix,  
$$
\Omega \equiv \Auu + ( \Vuu_1 + \Wuu_1 ) \Guu ( \Vuu_2 + \Wuu_2 )^* \ ,
$$
has the following Moore-Penrose generalized inverse:
$$
\Omega^+ = \Auu^+ \ - \
\Cuu_{2} \Vuu_2^* \Auu^+
\ - \Auu^+ \Vuu_1 \Cuu_{1}^* \ +
\Cuu_{2} ( \Guu^+ + \Vuu_2^* \Auu^+ \Vuu_1 ) \Cuu_1^*
.\eqno (2) $$

Proof: We recall that the Moore Penrose inverse is the unique
generalized inverse which satisfies the following four
conditions,(Ref. [5], p.26):
\vspace{.1in}

$(a) \ \Omega\Omega^+\Omega=\Omega, \ \
(b) \ \Omega^+\Omega\Omega^+ =\Omega^+ , \ $ 
\vspace{.1in}

$ (c) \ (\Omega\Omega^+)^* \ =\Omega\Omega^+, \ \ 
 (d) \ (\Omega^+\Omega)^* \ =\Omega^+ \Omega.$ 
\vspace{.1in}

The identity is verified by direct computation,
\vspace{.1in}

\noindent
${
\Omega\Omega^+ \equiv \Auu \Auu^+ \ - \
\Auu \Cuu_{2} \Vuu_2^* \Auu^+
\ - \Auu \Auu^+ \Vuu_1 \Cuu_{1}^* \ +
\Auu \Cuu_{2} ( \Guu^+ + \Vuu_2^* \Auu^+ \Vuu_1 ) \Cuu_1^*
}$ 
\vspace{.1in}

\noindent
$+ (\Vuu_1+\Wuu_1)\Guu (\Vuu_2+\Wuu_2)^* \Auu^+
- \ (\Vuu_1+\Wuu_1) \Guu (\Vuu_2+\Wuu_2)^* \Cuu_{2} \Vuu_2^* \Auu^+
$
\vspace{.1in}

\noindent
${ - ( \Vuu_1 + \Wuu_1 ) \Guu
\ (\Vuu_2+\Wuu_2)^*  \Auu^+ \Vuu_1 \Cuu_{1}^* }$
\vspace{.1in}

\noindent
${  +(\Vuu_1+\Wuu_1)\Guu (\Vuu_2+\Wuu_2)^* \Cuu_{2}
( \Vuu_2^* \Auu^+ \Vuu_1 ) \Cuu_1^*
} $
\vspace{.1in}

\noindent
${ +(\Vuu_1+\Wuu_1)\Guu (\Vuu_2+\Wuu_2)^* \Cuu_{2}  \Guu^+  \Cuu_1^* .} $

\bigskip
\noindent
Since $\Wuu_2$ is orthogonal to $\Auu^*$, we have
$\Auu\Wuu_2=0$ , $\Wuu^*_2 \Auu^+=0$ , and $\Vuu^*_2 \Wuu_2 = 0$,
which simplifies the previous expression to

\bigskip

$
\Omega \Omega^+ \equiv \Auu \Auu^+ \ 
 - \Auu \Auu^+ \Vuu_1 \Cuu_{1}^* \ +
 (\Vuu_1+\Wuu_1)\Guu \Vuu_2^* \Auu^+$
\vspace{.1in}

$- \ (\Vuu_1+\Wuu_1) \Guu \Wuu_2^* \Cuu_{2} \Vuu_2^* \Auu^+
- ( \Vuu_1 + \Wuu_1 ) \Guu
\ \Vuu_2^*  \Auu^+ \Vuu_1 \Cuu_{1}^* $
\vspace{.1in}

$ +
(\Vuu_1+\Wuu_1)\Guu \Wuu_2^* \Cuu_{2}  \Vuu_2^* \Auu^+ \Vuu_1\Cuu_1^*
+(\Vuu_1+\Wuu_1)\Guu \Wuu_2^* \Cuu_{2}\Guu^+\Cuu_1^* $.

\bigskip

This expression may be simplified using
$\Guu \Wuu_2^* \Cuu_2 \Guu^+ \Cuu_1^* = \Cuu_1^* $, and

\noindent
$\Guu \Wuu_2^* \Cuu_2 \Vuu_2^* = \Guu \Vuu_2^*$,
and  $\Auu\Auu^+\Vuu_1=\Vuu_1$ to
\vspace{.1in}

${
\Omega \Omega^+ \equiv \Auu \Auu^+  \ 
+ \Wuu_1 \Cuu_1^* },$
\vspace{.1in}

\noindent
and clearly condition (c) is satisfied. 

The corresponding identity for 
$\Omega^+ \Omega \equiv \Auu^+ \Auu  \ + \Cuu_{2}  \Wuu_2^* $
requires the decomposition to satisfy
 $\Auu^+ \Wuu_1=0$ , $\Wuu^*_1 \Auu=0$ , $\Vuu^*_1 \Wuu_1 = 0$,
and $\Vuu_2 \Auu^+ \Auu = \Vuu_2$. 
In addition, the matrix $\Guu$ must satisfy
$\Cuu_2 \Guu^+ \Cuu_1^*   \Wuu_1\Guu= \Cuu_2$ and 
$\Vuu_1  \Cuu_1^*   \Wuu_1 \Guu  = \Vuu_1 \Guu $.
These requirements guarantee that conditions (a), (b) and (d)
are also satisfied.
[]

\vspace{.1in}

Remark: The conditions that $\Guu$ and $\Wuu_i^* \Wuu_i$ have rank $k$ may
be replaced by the somewhat weaker but more complicated conditions
that  
$\Guu \Wuu_2^* \Cuu_2 \Guu^+ \Cuu_1^* = \Cuu_1^* $, 
$\Guu \Wuu_2^* \Cuu_2 \Vuu_2^* = \Guu \Vuu_2^*$,
$\Cuu_2 \Guu^+ \Cuu_1^*   \Wuu_1\Guu= \Cuu_2 $ and 
$\Vuu_1  \Cuu_1^*   \Wuu_1 \Guu  = \Vuu_1 \Guu $.


Note that the generalized inverse in (2)
is singular and tends to infinity as $\Wuu_i$ approaches to zero.
Thus (2) does not reduce to the (1) as the perturbation tends to zero.
When the perturbation of the column space of $\Auu$ is zero, i.e. 
$\Vuu \equiv 0$, theorem 1 simplifies to

$$
\Omega^+ = \Auu^+ \ + \ \Cuu_2 \Guu^+ \Cuu_1
.\eqno (3) $$

When $\Auu$ is a symmetric matrix, the column spaces of $\Auu$
and $\Auu^*$ are identical. Thus, for the case of symmetric $\Auu$
and $\Omega$, Thm. 1 reduces to  
\vspace{.1in}


{\bf Theorem 2}.  Let $\Auu$ be a symmetric $\ell \times \ell$ 
 matrix of rank $\ell_1$, $\ell_1 < \ell$,
 $\Vuu$ and $\Wuu$ be $\ell \times k$ matrices
and $\Guu$ be a $k \times k$ nonsingular matrix.
Let $\Vuu \in M(\Auu)$ and the columns of $\Wuu$ be orthogonal to $M(\Auu)$.
Let $\Buu \equiv \Wuu^* \Wuu$ have rank $k$. 
The matrix,  
$$
\Omega \equiv \Auu + ( \Vuu + \Wuu ) \Guu ( \Vuu + \Wuu )^* \ ,
$$
has the following Moore-Penrose generalized inverse:
$$
\Omega^+ = \Auu^+ \ - \
\Cuu \Vuu^* \Auu^+
\ - \Auu^+ \Vuu \Cuu^* \ +
\Cuu ( \Guu^+ + \Vuu^* \Auu^+ \Vuu ) \Cuu^*
.\eqno (4) $$

For concreteness, we specialise the preceding identities to the case of 
rank one perturbations. In this special case, $k \equiv 1$, and $\Vuu_i$
and $\Wuu_i$ reduce to $\ell$ vectors, $\vec{v_i}$ and $\vec{w_i}$.
In the nonsingular case, (1) reduces to Bartlett's 
identity [6]. It states for an arbitrary nonsingular
$\ell \times \ell$ matrix
$\Auu$ and $\ell$ vectors $\vec v_i$, 

$$
( \Auu + \vec{v_1} \vec{v_2}^* )^{-1}
= \Auu^{-1} \ - \
{ {( \Auu^{-1} \vec{v_1}) ( \vec{v_2}^* \Auu^{-1} ) \over
(1 + \vec{v_2}^* \Auu^{-1} \vec{v_1} ) }}
.\eqno (5)
$$

In this case, theorem 1 reduces to
the analogous result for an arbitrary singular
matrix $\Auu$ with a rank one perturbation which contains a component
perpendicular to the column space of $\Auu$. Noting that $\Guu \equiv 1$
and $\Cuu_i  \equiv \wv_i / |w_i|^2$, theorem 1 simplifies to the 
following result.

\vspace{.1in}

{\bf Theorem 3}.  Let $\Auu$ be a $\ell \times \ell$ 
matrix of rank $\ell_1$, $\ell_1 < \ell$, 
and $\vec{v_i}, \vec{w_i},\ i=1,2$ be $\ell$ vectors.
Let $\vv_1 \in M(\Auu)$ and $\vec{w_1}$ be orthogonal to $M(\Auu)$,
and $\vv_2 \in M(\Auu^*)$ and $\vec{w_2}$ be orthogonal to $M(\Auu^*)$.
Assume $\wv_2$ is parallel to $\wv_1$ and $\wv_i \ne 0$.  Let
$$
\Omega \equiv \Auu + ( \vec{v_1} + \vec{w_1} )( \vec{v_2} + \vec{w_2} )^* \ ,
$$
The Moore-Penrose generalized inverse is
$$
\Omega^+ = \Auu^+ \ - \
{ \vec{w_2} \vec{v_2}^* \Auu^+ \over |w_2|^2}  \ -
\ {\Auu^+ \vec{v_1} \vec{w_1}^*  \over |w_1|^2}
\ + \ (1 + \vec{v_2}^* \Auu^+ \vec{v} ) \
{\vec{w_2} \vec{w_1}^* \over |w_1|^2 |w_2|^2 } .\eqno (6)
$$

This generalized inverse
is singular and tends to infinity as $1 / |w_1| |w_2|$,
as $\vec{w_i}$ approaches to zero.  Thus (6) does not reduce to
Bartlett's identity.

The projection operator onto the row space of $\Omega$ is
$$
P_{X_T} = \Auu^+ \Auu \ + \ {\vec{w_i} \vec{w_i}^* \over |w_i|^2} \ .
$$

The symmetric version of Thm. 3 was originally developed and applied
by the author in his statistical analysis of magnetic fusion data [7]. 
To estimate the regression parameters in ordinary least squares
regression, the sum of squares and products (SSP) matrix
needs to be inverted. We  apply Thm. 3 to determine the inverse
of the SSP matrix in terms of the inverse of the 
covariance matrix of the covariates.

We decompose the independent
variable vector, $\vec{x}$ into a mean value vector,
$\vec{\bar x}$ and a fluctuating part, $\vec{\tilde x}$.
Thus the $i$-th individual observation has the form
$$
\vec{x}_i =  \vec{\bar x} + \vec{\tilde x}_i
\eqno .$$
Let $\Xuu$
denote the $n \times \ell$ data matrix whose rows consist of 
$\xv_i^*$ and $\tilde{\bf X} $ be the centered data matrix
whose rows consist of $\vec{\tilde x}_i^*$.

We assume that some of the independent variables,
$x_k$, have not been varied. Thus $\tilde{\bf X}^* \tilde{\bf X}$
is singular. The inverse of the uncentered sum of squares and
crossproducts matrix, ${\bf X}^*{\bf X}$ can now be expressed in
terms of the Moore Penrose generalized inverse of the centered covariance matrix,
$\tilde{\bf X}^* \tilde{\bf X}$.

We decompose a multiple of the mean value vector, $\sqrt{n} \xbv$, into
$\vv + \wv$, where $\vv \in M( \tilde{\bf X}^* \tilde{\bf X})$
and $\wv \perp M(\tilde{\bf X}^* \tilde{\bf X})$.

The data matrix has the form
$$
{ \bf X^* X} = \tilde{\bf X}^* \tilde{\bf X} + n
\vec{\bar x} \vec{\bar x}^{T} =
\tilde{\bf X}^* \tilde{\bf X} + (\vv + \wv)(\vv + \wv)^*
.$$

Thus we have rewritten ${\bf X^* X}$ in a form appropriate to the application
of theorem 3. 

In conclusion, the application of these 
matrix identities requires the decomposition
of $\Xuu_i$ into the orthogonal components, $\Vuu_i$ and $\Wuu_i$.
Thus our theorems are most useful in situations where the decomposition
is trivial.

\noindent
{\it Acknowledgments}

The helpful comments of the referees  are gratefully acknowledged.

\bigskip

\begin{center}
{\bf REFERENCES}
\end{center}
\begin{enumerate}
\item W.J. Duncan, ``Some devices for the solution of large sets of
simultaneous equations (with an appendix on the reciprocation of 
partitioned matrices)",
{\it The London, Edinburgh and Dublin Philosophical
Magazine and Journal of Science}, Seventh Series, {\bf 35}, p. 660, (1944).
\item H.V. Henderson and S.R. Searle, 
``On deriving the inverse of a sum of matrices",
SIAM Review, {\bf 23}, p.53, (1981).
\item D.V. Ouellete, ``Schur complements and statistics",
Journal of Linear Algebra, {\bf 36}, p. 187, (1981).
\item W.W. Hager, ``Updating the inverse of a matrix",
SIAM Review, {\bf 31}, p.221, (1989).
\item C.R. Rao, {\it Linear Statistical Inference and Its Applications},
p. 26,33, J. Wiley and Sons, New York, 1973.
\item M.S. Bartlett, ``An inverse matrix adjustment arising in
discriminant analysis",
{\it Annals of Mathematical Statistics}, vol. 22, p107,
(1951).
\item K.S. Riedel, 
New York University Report MF-118,
National Technical Information Service document no. DOE/ER/53223-98
"Advanced Statistics for Tokamak Transport: Collinearity and
Tokamak to Tokamak Variation" (1989).
\end{enumerate}
\newpage
\end{document}